\documentclass{sig-alternate-10pt}

\usepackage[margin=1in]{geometry}

\usepackage{times}
\usepackage{amsmath}
\usepackage{amssymb}
\usepackage{graphicx}
\usepackage{booktabs}
\usepackage{hyperref}
\usepackage{xcolor}
\usepackage{cite}
\usepackage[normalem]{ulem}
\usepackage{float}
\usepackage[most]{tcolorbox}

\usepackage{titlesec}
\titleformat{\section}
{\normalfont\large\bfseries}
{\thesection.}{0.5em}{}
\titlespacing{\section}{0pt}{12pt}{6pt}

\begin{document}
	\pagestyle{empty}
	
	\title{Not Every Dependency Is Worth Discovering:\\
		Toward Value-Driven Data Dependency Discovery}
	
	\author{
		\alignauthor
		Xiaolong Wan, Xixian Han\\
		\affaddr{Harbin Institute of Technology, China}\\
		\email{wxl@hit.edu.cn, hanxx@hit.edu.cn}
	}	

	\maketitle
	
	\begin{abstract}
		Data dependency discovery has traditionally focused on identifying dependencies that hold in the data or are statistically strong. Yet a dependency may be valid without being valuable: it may be irrelevant to the governance task, redundant given existing knowledge, or too costly to discover, validate, maintain, and apply. We call for a shift from validity-driven to value-driven dependency discovery. We define dependency use value decision-theoretically as the expected reduction in task-specific loss from incorporating a dependency into the governance process, and define net value by further accounting for lifecycle costs. Building on this framework, we outline principles for value-aware search, validation, dependency-set selection, and maintenance, and identify a research agenda spanning value estimation before full discovery, loss and cost learning, budgeted set selection, lifecycle monitoring, and benchmarking.
	\end{abstract}

	\section{The Missing Question}
	
	Data dependency discovery has long focused on a single question~\cite{DBLP:series/synthesis/2018Abedjan,SongChen2023}: which dependencies hold in the data? Yet a different question has received far less attention: which of those dependencies are actually worth discovering? \emph{A dependency may be valid without being valuable}.
	
	Dependency discovery has traditionally been defined in terms of the data alone: find the dependencies that hold~\cite{DBLP:journals/cj/HuhtalaKPT99,	DBLP:journals/pacmmod/BleifussPBSN24, DBLP:journals/tkde/WanHWL24}, or those that are statistically strong~\cite{DBLP:journals/vldb/ParciakWHNPV25}. Even top-$k$ approaches usually prioritize dependencies using criteria such as approximate strength, redundancy, subjective relevance, or diversity~\cite{DBLP:journals/pacmmod/FanHXZ24,	DBLP:journals/pacmmod/FanHWX23, DBLP:journals/corr/abs-2605-24925,	DBLP:journals/corr/abs-2601-10130}. Despite these differences, these approaches share the same problem boundary: dependency discovery is considered complete once the dependencies in the data have been identified, measured, or prioritized. Whether those dependencies improve a governance decision, and whether their benefits justify the costs of discovering and using them, remains outside the problem definition.
	
	We argue that defining dependency discovery in this way is too narrow. The value of a dependency is not determined by the data alone, nor is it an intrinsic property of the dependency itself. It depends on the governance task at hand, the decisions the dependency enables, and the costs of discovering, validating, maintaining, and applying it~\cite{Wilson2015}. The same dependency may be highly valuable for one task and worthless for another. Dependency value is therefore contextual: \emph{it depends on the data, the task, and the consequences of using the dependency}.
	
	\begin{figure*}
		\centering
		\includegraphics[scale = 0.6]{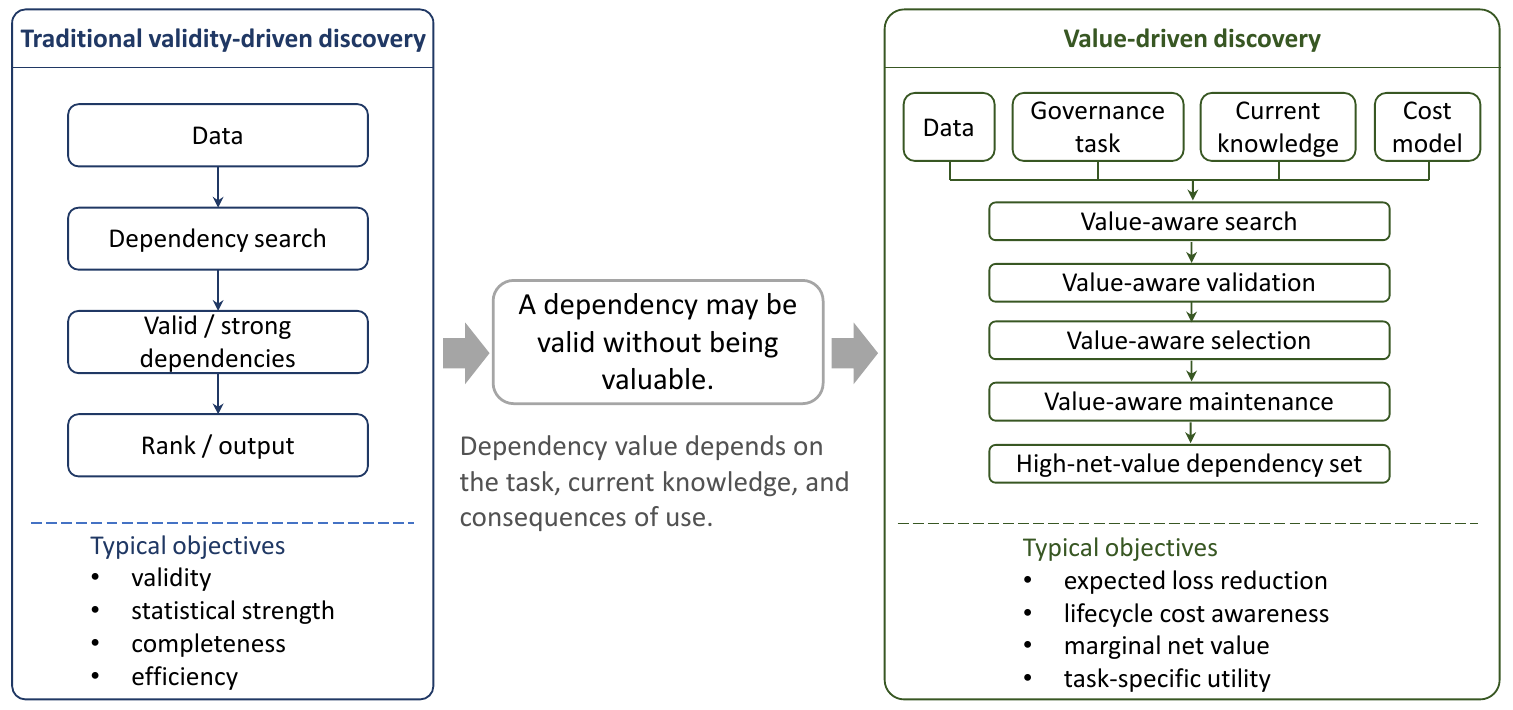}
		\caption{From validity-driven to value-driven dependency discovery. Value-driven discovery incorporates the governance task, current knowledge, and lifecycle costs, and uses estimated marginal net value to guide search, validation, selection, and maintenance.}
		\label{fig:paradigm}
	\end{figure*}	

	We therefore call for a shift from validity-driven to value-driven dependency discovery, as illustrated in Figure~\ref{fig:paradigm}. This is not a call for better post-hoc ranking of discovered dependencies. Value must shape what is searched for, what is validated, what is returned, and what is maintained. It raises a new set of questions: how should dependency value be defined, how can it be estimated efficiently, and how should discovery algorithms change when value becomes the objective?
	
	\section{Why Valid Is Not Valuable}
	
	Validity and value answer different questions. A dependency is valid if it holds in the data---exactly, or within some prescribed error tolerance~\cite{DBLP:series/synthesis/2018Abedjan,SongChen2023}. It is valuable if knowing and using it improves a governance decision. The first is a statement about the data; the second is a statement about consequences. A dependency can be perfectly valid and yet make no difference to any decision a governance process would take.

	Validity can fail to imply value for several reasons. A dependency may duplicate what is already known, adding no decision-relevant information and leaving the optimal governance action unchanged. It may involve attributes irrelevant to the task, so that no governance action depends on it regardless of whether the dependency holds. Even when a dependency is task-relevant, using it may produce harms and costs that offset its benefits: false alerts that require human review, incorrect actions triggered by misleading violations, or maintenance overhead that grows as the data evolves~\cite{DBLP:books/acm/IlyasC19}. What ultimately matters is not merely whether a dependency holds, but whether acting on it improves the outcome~\cite{DBLP:journals/pvldb/KrishnanWWFG16}.
	
	The same dependency can have different value under different tasks. A dependency useful for error detection may be uninformative for schema normalization, and one that guides repair effectively may produce excessive alerts when used for monitoring. There is no single, task-independent notion of dependency value. What counts as valuable depends on the task, the consequences it assigns to different outcomes, and the costs it associates with discovery and use. Defining dependency value precisely, and in a way that is sensitive to task and cost, is the problem we address next.
	
	\section{Decision-Theoretic Dependency Value} \label{sec:framework}
	
	To define dependency value precisely, we need a small set of primitives. A \emph{governance task} $T$ specifies what the dependency will be used for---error detection, data repair, schema normalization, or another objective. A \emph{knowledge state} $K$ represents what is currently known: the dependencies already discovered and validated, along with any prior information about data quality. A \emph{set of governance actions} $\mathcal{A}$ contains the actions available to the process---which tuples to flag, which values to repair, or which constraints to enforce. A \emph{loss function} $L(a, \theta)$ assigns a cost to taking action $a \in \mathcal{A}$ when the underlying state relevant to the task is $\theta$. Together, these primitives ground the notion of value in the consequences of governance decisions, rather than in data-intrinsic properties alone~\cite{DBLP:journals/tssc/Howard66,Wilson2015}. The framework is normative: it is intended for governance tasks that can be represented by an operational action space, a meaningful loss model, and at least approximate estimates of lifecycle costs.
	
	Given these primitives, the \emph{governance use value} of a dependency $\phi$ is the expected reduction in loss that results from incorporating $\phi$ into the knowledge state. Let $K'=\mathcal{U}(K,\phi)$ denote the knowledge state obtained by incorporating $\phi$ into $K$; let $a^*(K)$ denote the action minimizing expected loss under $K$, and let $P(\theta\mid K)$ denote the posterior distribution over $\theta$ given $K$, reflecting uncertainty about the underlying state relevant to the task. Then:
	\begin{align}
		V_{\mathrm{use}}(\phi\mid T,K)
		&=
		\mathbb{E}_{\theta\sim P(\theta\mid K)}\notag\\
		&\quad
		\bigl[
		L(a^*(K),\theta)
		-
		L(a^*(K'),\theta)
		\bigr].
	\end{align}	
	The action $a^*(K')$ is selected using the updated knowledge state, while both actions are evaluated from the perspective of $K$, making their consequences comparable. This is a marginal use value conditional on incorporating $\phi$: it measures how much the best available action is expected to improve if $\phi$ is added to $K$. A dependency has positive use value only if the updated knowledge state leads to an action with lower expected loss. If $\phi$ is redundant given $K$, or irrelevant to $T$, then $a^*(K') = a^*(K)$ and $V_{\mathrm{use}} = 0$.
	
	Use value alone does not determine whether a dependency is worth discovering. Acquiring, validating, maintaining, and applying a dependency all carry costs. We define the \emph{governance net value} as:
	\begin{align}
		V_{\mathrm{net}}(\phi \mid T, K) &= V_{\mathrm{use}}(\phi \mid T, K) \notag\\
		&\quad - C_{\mathrm{disc}}(\phi) - C_{\mathrm{val}}(\phi) \notag\\
		&\quad - C_{\mathrm{maint}}(\phi) - C_{\mathrm{apply}}(\phi)
	\end{align}	
	where $C_{\mathrm{disc}}$ is the cost of searching for and identifying $\phi$, $C_{\mathrm{val}}$ is the cost of confirming that it holds, $C_{\mathrm{maint}}$ is the cost of monitoring its validity as data evolves, and $C_{\mathrm{apply}}$ is the cost of integrating it into the governance process. A dependency is worth acquiring only if $V_{\mathrm{net}} > 0$: even a dependency with high use value may not justify the effort if these costs are too high.
		
	To make these definitions concrete, consider error detection as the governance task~\cite{DBLP:journals/ftdb/IlyasC15}. The action space consists of decisions about which tuples to flag for review. The loss function captures the consequences of those decisions:
	\begin{align}
		L_{\mathrm{det}}(a, \theta) &= c_{\mathrm{FN}} \cdot FN(a, \theta) \notag\\
		&\quad + c_{\mathrm{FP}} \cdot FP(a, \theta) 
		+ c_{V} \cdot N_{\mathrm{check}}(a)
	\end{align}
	where $c_{\mathrm{FN}}$ and $c_{\mathrm{FP}}$ are the costs of missed errors and false alerts respectively, and $c_V$ is the per-tuple verification cost. The relative magnitude of these costs varies across domains: in a safety-critical setting, $c_{\mathrm{FN}}$ may dominate, whereas in a high-throughput pipeline with expensive human review, $c_V$ may become the binding constraint. $FP$ and $N_{\mathrm{check}}$ are not double-counted: $FP$ captures the decision cost of incorrectly flagging a clean tuple, while $N_{\mathrm{check}}$ captures the resource cost of reviewing any flagged tuple, regardless of whether the flag is correct. Under this instantiation, $V_{\mathrm{use}}$ measures the expected reduction in detection loss when $\phi$ is incorporated into the detection process. A dependency has positive use value for detection when incorporating it leads to an action whose expected reduction in missed-error loss exceeds the additional costs of false alerts and verification. A dependency is worth discovering for this task only if this benefit survives the deduction of discovery, validation, maintenance, and application costs. For data repair, loss captures the cost of incorrect modifications, missed fixes, and potential semantic distortion; for schema normalization, it captures redundancy overhead and update anomaly costs~\cite{DBLP:journals/vldb/ZhangL25}. The framework accommodates all of these; the loss function is the point where task semantics enter.
	
	\begin{figure*}
		\centering
		\includegraphics[scale = 0.6]{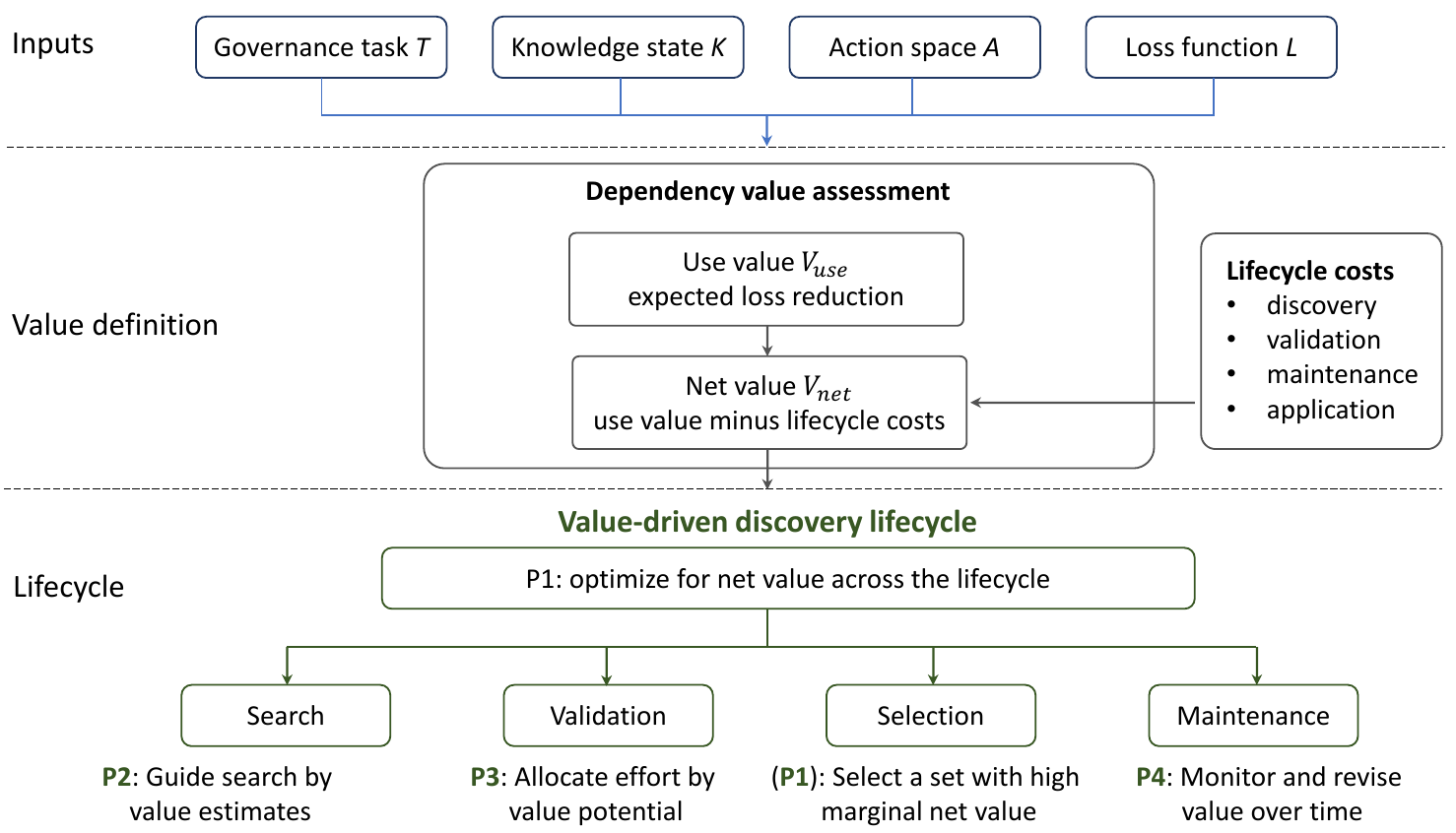}
		\caption{Decision-theoretic dependency value across the discovery lifecycle. Governance task, current knowledge, available actions, and task-specific losses jointly determine use value $V_{\mathrm{use}}$ as expected loss reduction. Net value $V_{\mathrm{net}}$ further deducts lifecycle costs and guides search, validation, selection, and maintenance through the principles in Section~\ref{sec:principles}.}
		\label{fig:value-framework}
	\end{figure*}
	
	\section{Principles for Value-Driven Discovery} \label{sec:principles}
	
	Adopting net value as the optimization target is not merely a change of metric; it requires rethinking how dependencies are searched for, validated, selected, and maintained. We outline four principles for carrying this objective through the dependency discovery lifecycle, as illustrated in Figure~\ref{fig:value-framework}.
	
	\medskip
	\noindent
	\textbf{P1.} Optimize for net value, not validity or
	statistical strength.\\[2pt]
	\textbf{P2.} Let value estimates guide search, not merely
	rank results.\\[2pt]
	\textbf{P3.} Allocate validation effort according to value
	potential.\\[2pt]
	\textbf{P4.} Monitor and revise dependency value over time.
	\medskip
	
	\emph{\uline{P1}: Optimize for net value, not validity or statistical strength}. The goal of dependency discovery should be to maximize governance net value, rather than to enumerate valid dependencies or rank them by data-intrinsic measures. A dependency is worth discovering only if the governance benefit it enables exceeds the full cost of acquiring, validating, maintaining, and applying it. Since dependencies may be redundant or complementary, the relevant quantity is marginal net value: how much a dependency adds, given what is already known, after accounting for its costs. Under a resource budget, the objective becomes identifying a set $S$ that maximizes $V_{\mathrm{net}}(S\mid T,K)$ subject to resource constraints, rather than scoring dependencies independently.	

	\emph{\uline{P2}: Let value estimates guide search, not merely rank results}. Value-driven discovery is not a post-hoc step that reorders an already completed search. Estimated value bounds should guide which candidates are explored first, which are pruned early, and when further search is unlikely to justify its cost. This requires computable proxies for $V_{\mathrm{net}}$ that can be evaluated cheaply during search, before full validation. Value proxies must themselves be lightweight: their evaluation cost must represent a negligible fraction of the discovery cost they guide, or value-aware search becomes a source of negative net value in its own right.
		
	\emph{\uline{P3}: Allocate validation effort according to value potential}. Not all candidates deserve equal validation resources. A candidate with high value potential justifies more thorough validation, whereas one whose value upper bound falls below the minimum net value needed to justify further resource allocation can be pruned without full validation. Validation should therefore be viewed as a sequential decision: each additional check should be chosen for its expected reduction in uncertainty about whether the candidate is worth selecting. Additional validation is justified only when its expected contribution to a better discovery decision exceeds its cost.
	
	\emph{\uline{P4}: Monitor and revise dependency value over time}. A dependency that is valuable today may not remain so. Data distributions shift, tasks change, costs evolve, and new dependencies may render existing ones redundant. Validity may remain stable while value changes. Dependencies should therefore be reassessed over time, and revalidated, reprioritized, or retired as their marginal net value changes.
	
	\section{Research Agenda}
	
	The principles outlined in Section~\ref{sec:principles} raise five concrete research problems. While decision theory~\cite{DBLP:journals/tssc/Howard66, Wilson2015}, cost-sensitive learning~\cite{DBLP:conf/ijcai/Elkan01}, and submodular optimization~\cite{DBLP:conf/aaai/KrauseG07} offer partial foundations, realizing value-driven discovery requires these to be developed and integrated in ways that do not yet exist for dependency discovery specifically. The five problems are not independent: R1 and R2 are tightly coupled, because value estimation depends on the loss and cost model, while learning useful losses and costs depends on how value is defined for discovery decisions. Together, R1 and R2 enable value-aware search and pruning. R3 extends single-dependency estimates to budgeted set selection. R4 addresses value change over time. R5 provides the evaluation infrastructure needed to measure progress on all four. These problems define a research path toward value-driven discovery.
		
	\begin{figure}[t]
		\begin{tcolorbox}[colback=white, colframe=black, arc=3pt, boxrule=0.5pt]
			\textbf{R1.} Estimating value without full discovery.\\[2pt]
			\textbf{R2.} Learning task-specific losses and costs.\\[2pt]
			\textbf{R3.} Selecting dependency sets under budget constraints.\\[2pt]
			\textbf{R4.} Monitoring value across the dependency lifecycle.\\[2pt]
			\textbf{R5.} Benchmarking value-driven discovery.
		\end{tcolorbox}
	\end{figure}
	
	\emph{\uline{R1}: Estimating value without full discovery.} The central tension in value-driven discovery is that accurate value estimation typically requires a dependency to be found and validated, yet value estimates are needed to guide the search before that point. Resolving this tension requires computable proxies or bounds for $V_{\mathrm{net}}$ that can be evaluated from partial evidence, including attribute statistics, sampled partitions, or intermediate search results, without completing full discovery and validation. A key question is whether existing data-intrinsic measures can serve as reliable proxies under identifiable conditions, and where they fail.
	
	\emph{\uline{R2}: Learning task-specific losses and costs.} The framework in Section~\ref{sec:framework} assumes that the loss function $L$ and the cost terms are known. In practice, they rarely are. The relative cost of a missed error versus a false alert depends on the application domain, the downstream use of the data, and organizational priorities that are often implicit. Similarly, discovery, validation, maintenance, and application costs depend on data scale, schema complexity, and available infrastructure. Research is needed on how to elicit, learn, or approximate these quantities from historical governance records, user feedback, or domain knowledge, and on how sensitive discovery decisions are to errors in the specified loss and cost models. Precise cost values may not be necessary in all settings. Determining whether ordinal cost relationships or bounded cost intervals suffice to preserve search priorities and support safe pruning is an open problem whose resolution would substantially reduce the practical burden of value-driven discovery.
	
	\emph{\uline{R3}: Selecting dependency sets under budget constraints.} Individual dependency values do not determine which set of dependencies should be discovered and deployed. Dependencies may be redundant---two dependencies that are each valuable alone may together add little beyond what either contributes individually---or complementary, so that neither is sufficiently valuable in isolation but both together are. The selection problem is to identify a set $S$ that maximizes $V_{\mathrm{net}}(S \mid T, K)$ subject to a resource budget. This connects to submodular optimization and budgeted observation selection~\cite{DBLP:conf/aaai/KrauseG07}, but the structure of dependency value functions and whether they admit tractable approximations remain open questions.
	
	\emph{\uline{R4}: Monitoring value across the dependency lifecycle.} A dependency that is valuable when discovered and validated may gain or lose value as data evolves, tasks shift, costs change, or new dependencies are found. Existing incremental methods maintain dependency validity under data updates~\cite{DBLP:conf/icde/XiaoYTMW22, DBLP:journals/corr/abs-2601-16025}, but value may change even when validity does not. The challenge is not only to detect when a dependency has become invalid, but also to determine when its marginal net value no longer justifies continued maintenance and use. This requires methods for tracking value changes incrementally, triggering revalidation selectively, and deciding when to reprioritize, retire, or replace dependencies, without rediscovering the entire dependency structure from scratch.
	
	\emph{\uline{R5}: Benchmarking value-driven discovery.} Existing benchmarks for dependency discovery measure correctness, completeness, runtime, and memory~\cite{DBLP:journals/pvldb/PapenbrockEMNRZ15}, reflecting the goals of validity-driven discovery. These metrics are insufficient for evaluating value-driven systems, which should be assessed by the governance benefit they deliver relative to the resources they consume. New evaluation protocols are needed that specify governance	tasks, cost parameters, and ground-truth task outcomes, and that measure both value-estimation quality and task-level utility rather than only dependency-level measures such as validity, strength, or redundancy. Without such benchmarks, progress in value-driven discovery will be difficult to demonstrate and compare.
	
	\section{Positioning Value-Driven Discovery}
	
	Existing criteria for ranking or selecting dependencies characterize either data-intrinsic properties---such as validity, approximate strength, or redundancy---or user-specified notions of interest, such as subjective relevance, as well as properties of the selected set, such as diversity~\cite{DBLP:journals/pacmmod/FanHWX23,		DBLP:journals/pacmmod/FanHXZ24,	DBLP:journals/corr/abs-2605-24925,		DBLP:journals/corr/abs-2601-10130}. These criteria are useful for prioritizing dependencies and, in some cases, for guiding search and pruning. However, they do not directly quantify the consequences of incorporating a dependency into a governance decision. Redundancy is not loss reduction, relevance is not decision improvement, and diversity is not marginal net value. Value-driven discovery therefore differs not merely in when a criterion is applied, but in what the criterion represents: it estimates how much a dependency improves a governance decision, after accounting for the full costs of discovering, validating, maintaining, and applying it.
	
	Task-sensitive data quality assessment offers a related	contextual perspective. Recent work has argued that assessing data quality dimensions such as accuracy, completeness, and relevancy depends on the data, the source, the system, the task, and the humans involved~\cite{DBLP:journals/sigmod/MohammedEHNS25}. We share the view that quality judgments are task-dependent, but ask a different question: not how to assess data quality under a given task, but whether acquiring and using a particular dependency improves the governance decision associated with that task. Their framework contextualizes quality assessment, while ours contextualizes the governance value of the dependencies used to support such decisions.
	
	Value-of-information (VoI) theory provides the decision-theoretic foundation for asking whether information is worth acquiring~\cite{DBLP:journals/tssc/Howard66,Wilson2015}. The core idea---that the value of information lies in its ability to improve a decision, rather than in its intrinsic properties---directly motivates our framework. However, general VoI theory does not by itself address the specific challenges of dependency discovery: searching a combinatorial candidate space, validating dependencies at nontrivial cost, selecting sets under redundancy and complementarity, and tracking and revising dependency value as data and tasks evolve. We instantiate  VoI reasoning for this setting through $V_{\mathrm{use}}$, $V_{\mathrm{net}}$, and a lifecycle of value-aware search, validation, selection, and maintenance.
		
	\section{Conclusion}
	Data dependency discovery has long treated validity as its primary objective. We have argued that validity is necessary but not sufficient: a dependency that holds in the data may still be irrelevant to the governance task, redundant given existing knowledge, or not worth the cost of acquiring and maintaining it. The question of whether a dependency is worth discovering has received far less attention. We have proposed a decision-theoretic framework that defines use value as expected task-specific loss reduction and net value by further accounting for lifecycle costs. Building on this framework, we have outlined four principles for allowing net value to shape the discovery lifecycle, and identified five open problems whose resolution is necessary to realize value-driven discovery in practice. The broader implication extends beyond dependency discovery: any data management process that produces knowledge objects---rules, constraints, patterns, or metadata---faces the same question: is this knowledge worth acquiring and using? We hope this work	encourages the community to ask not only what holds, but what is worth knowing and acting upon.

	

\end{document}